\documentclass[]{raa}                  % preprint format
\usepackage{graphicx,times}             %for PS/EPS graphics inclusion, new
\input{epsf.sty}                        %for PS/EPS graphics inclusion, old
\input{psfig.sty}                       %for PS/EPS graphics inclusion, old

\begin{document}

   \title{Preferred alignments of angular momentum vectors of galaxies in six dynamically unstable Abell clusters
%\,$^*$
%\footnotetext{$*$ Supported by the National Natural Science Foundation of China.}
}
%   \subtitle{I. Place Your Subtitle Here}
   \volnopage{Vol. 000 No.0, 000--000}      %%preserved for Editor. DOn't remove!
   \setcounter{page}{1}          %%starting page, preserved for Editor. DOn't remove!

   \author{S. N. Yadav
      \inst{1}
   \and B. Aryal
      \inst{1}
   \and W. Saurer
      \inst{2}
   }
%% Here is an example of three authors come from different institutes.
%% For single author or all the authors from an institute, use "\inst{}" only

   \institute{Central Department of Physics, Tribhuvan University, Kirtipur, Nepal {\it ysibnarayan@yahoo.com}\\
%% Please give the E-mail address of the author, to whom future correspondence and
%% offprint requests will be sent.
        \and
             Central Department of Physics, Tribhuvan University, Kirtipur, Nepal\\
        \and
             Institute of Astro-particle Physics, Innsbruck University, A-6020 Innsbruck, Austria\\
   }

\date{Received~~2016 October 16; accepted~~2017~~Feb 22}

\abstract{A spatial orientation of angular momentum vectors of
galaxies in six dynamically unstable Abell clusters (S1171, S0001,
A1035, A1373, A1474 and A4053) is studied. For this,
two-dimensional observed parameters (e.g., positions, diameters,
position angles) are converted into three-dimensional rotation
axes of the galaxy using `position angle - inclination' method.
The expected isotropic distribution curves for angular momentum
vectors are obtained by performing random simulations. The
observed and expected distributions are compared using several
statistical tests. No preferred alignments of angular momentum
vectors of galaxies are noticed in all six dynamically unstable
clusters supporting hierarchy model of galaxy formation. These
clusters have a larger value of velocity dispersion. However,
local effects are noticed in the clusters that have substructures
in the 1D-3D number density maps.  }

   \authorrunning{S. N. Yadav, B. Aryal \& W. Saurer }            %author_head in even pages
   \titlerunning{Preferred alignments of galaxies in six clusters}  % title_head in odd pages

   \maketitle
%% The author head (on even pages) and the title head (on odd pages) will be
%% automatically extracted from \author{} and \title{}. Whenever the title is too long,
%% you will be asked to supply a shorter one by inserting either \authorrunning{} or
%% \titlerunning{} before \maketitle. Anyway, you can specify your own heads.
%%
%%
%% Note: In the following text body of your manuscript, please note several differences from
%%       other major journals:
%% (1) \subsection{Please Capitalize the First Letter of Each Notional Word in Subsection Title}
%% (2) Please Capitalize the First Letter of Each Notional Word in all tables' captions

%
%________________________________________________ sections below
%
\section{Introduction}           %% first-level sections will be auto-capitalized
\label{sect:intro}

The formation of galaxy cluster is one of the major unsolved
problems of modern astrophysics. The process by which larger
structures (e.g., clusters, superclusters) are formed through the
continuous merging of smaller structures (e.g., galaxy, galaxy
groups) is called hierarchical clustering, which is supported by
concordance model ($\Lambda$CDM). The study of preferred alignment
of angular momentum vectors of galaxies in the clusters is one of
the most effective ways of testing the concordance model.
Godlowski et al. (2003) described Li (1988) model and showed the
relation between the angular momenta and the masses of the large
scale structures. This relationship was observationally tested by
several authors (Godlowski et al. 2005, Hu et al. 2006, Aryal \&
Saurer 2006, Aryal et al. 2007, 2008, Godlowski et al. 2010,
Godlowski 2012) and found the vanishing angular momenta for less
massive structures and non-vanishing for larger structures.

Aryal et al. (2013) studied preferred alignments of angular
momentum vectors of galaxies in six rotating clusters (A954,
A1139, A1399, A2162, A2169, and A2366) that are dynamically stable
and have a single peak in 1D-3D number density maps. These
clusters have no substructures. They found a random orientation of
angular momentum vectors of galaxies in all six clusters,
supporting hierarchy model (Peebles 1969).

In the present work we intend to study preferred alignments of
angular momentum vectors of galaxies in six Abell clusters namely
S1171, S0001, A1035, A1373, A1474 and A4053 that have multiple
peaks in 1D-3D number density maps with a larger value of velocity
dispersion. These clusters have substructures. We intend to find
out the answer of the following: (1) does the orientation of
angular momentum vectors of galaxies that have substructures favor
hierarchical clustering? (2) do the clusters with large velocity
dispersion prefer a random orientation of angular momentum vectors
of galaxies? and finally (3) does the substructure formation cause
large velocity dispersion in the cluster? Our aim is to compare
results with concordance model. The database and the methods are
described in sections 2 and 3. Our results and conclusions are
presented in sections 4 and 5.

\section{Database}
\label{sect:Obs}

Hwang \& Lee (2007, HL hereafter) proposed six rotating clusters
(A954, A1139, A1399, A2162, A2169, and A2366) that are in
dynamical equilibrium and show a single peak in 1D-3D number
density maps. In addition, six dynamically unstable clusters
(S1171, S0001, A1035, A1373, A1474 and A4053) that have multiple
number-density peaks in 1D-3D maps with a large velocity
dispersion are presented. After investigating substructure using
Dressler-Scectman, HL classified rotating clusters into two
category: (1) clusters with single number density peak and hence
are in dynamical equilibrium and (2) clusters with multiple
number-density peak and are dynamically unstable. In both cases,
clusters have a very large value of velocity dispersion. In this
paper we study the preferred alignments of angular momentum
vectors of galaxies in the Abell clusters S1171, S0001, A1035,
A1373, A1474 and A4053. Table 1 lists the the database (positions,
BM type classification, mean redshift, velocity dispersion, number
of galaxies in the cluster and its morphology) of six clusters
(HL, Hwang 2011) used for this study.

\begin{table*}
\caption[]{Database of six clusters that have multiple
number-density peaks in 1D-3D maps (HL). The first column lists
the Abell name followed by their positions (Abell et al. 1989).
The fourth-eighth columns give BM type cluster morphology (Bautz
\& Morgan 1970), mean radial velocity ($\overline{cz}$), velocity
dispersion ($\sigma_{\rm v}$), number (N) of galaxies in the
cluster and its morphology, as given in HL.}
$$
 \begin{array}{p{0.07\linewidth}ccccccccc}
            \hline
            \noalign{\smallskip}
            Abell  &  $R.A.$     & $Dec.$     &  $BM-type$ &  $$\overline{cz}$$  &  $$\sigma_{\rm v}$$    &  $N$  & $morphology$ \\
                     &  $(J2000)$  & $(J2000)$  &            &  $(km s$^{-1})$$  &  $(km s$^{-1})$$ &  & \\
                   \noalign{\smallskip}
            \hline
            \noalign{\smallskip}
            S1171      &   00^{\rm h}01^{\rm m}21.70^{\rm s}   &   27^{\circ}32'18.0'' &   $II$      &   8377    &   646   &   42 &  $Spherical$   \\
            S0001      &   00^{\rm h}02^{\rm m}33.93^{\rm s}   &   30^{\circ}44'06.2'' &   $I$       &   8815    &   577   &   51 &  $Elongated$   \\
            A1035      &   10^{\rm h}32^{\rm m}14.16^{\rm s}   &   40^{\circ}14'49.2'' &   $II-III$  &   21753   &   1825  &   97 &  $Spherical$   \\
            A1373      &   11^{\rm h}45^{\rm m}30.95^{\rm s}   &   02^{\circ}27'12.9'' &   $III$     &   37595   &   1768  &   48 &  $Elongated$   \\
            A1474      &   12^{\rm h}07^{\rm m}57.20^{\rm s}   &   14^{\circ}57'18.0'' &   $III$     &   24151   &   714   &   60 &  $Elongated$   \\
            A4053      &   23^{\rm h}54^{\rm m}45.39^{\rm s}   &   27^{\circ}40'52.8'' &   $III$     &   20691   &   1366  &   76 &  $Elongated$   \\
            \noalign{\smallskip}
            \hline
         \end{array}
     $$
\end{table*}

%__________________________________________________________________
\section{Method}

The PA-inclination method is used to convert two dimensional given
parameters (positions, diameters, position angles) into three
dimensional (galaxy rotation axes: angular momentum vectors and
its projections) (Flin \& Godlowski 1986). The expected isotropic
distribution curves for angular momentum vectors and its
projections are determined by performing a random simulation
(Aryal \& Saurer 2000). The observed distributions are compared
with the expected using various statistical tests.

\subsection{Observed distribution: angular momentum vectors of galaxies}

The angular momentum vectors ($\theta$) of galaxies and its
projections ($\phi$) to the galactic (G) and supergalactic (S)
planes are obtained by using the method described by Flin \&
Godlowski (1986). For this, SDSS/2dFGRS database (positions,
position angles and diameters) provided by Hwang (2011) are used.
In the previous works (Godlowski 1994, Baier et al. 2003, Hu et
al. 2006, and the references therein), authors have studied the
preferred alignments of galaxies in clusters with respect to
galactic or supergalactic (or both) system. The formulae to obtain
angular momentum vectors ($\theta$) and its projection ($\phi$) in
S-system are as follows (Flin \& Godlowski, 1986):
%____________________________________________________________
\begin{equation}
\sin \theta = - \cos i \sin B \pm \sin i \sin P \cos B
\end{equation}
%_____________________________________________________________
%_____________________________________________________________
\begin{equation}
\sin \phi = (\cos \theta)^{-1} [-\cos i\cos B \sin L + \sin i (\mp
\sin P\sin B\sin L \mp \cos P\cos L).
\end{equation}
%_____________________________________________________________
The inclination angle ($i$) is the angle between line-of-sight and
the normal to the plane of the galaxy. This angle can be
calculated using Holmberg's (1946) formula: cos$^2i$ =
[$(b/a)^2$--$q$$^2$]/(1--$q$$^2$) where $b/a$ is the measured
axial ratio and $q$ is the intrinsic flatness of disk galaxies.
Since the morphological information of galaxies are not known,
therefore the value of intrinsic flatness ($q$) is assumed to be
0.2 (Aryal et al. 2007, Godlowski 2011a). The parameters $L$, $B$
and $P$ are the supergalactic longitude, latitude and position
angle, respectively.

The formulae (1) and (2) give two solutions for a galaxy. The
reason for this is the approaching and receding sides of a galaxy,
which can not be identified in our database. Therefore, there are
four solutions for the preferred alignments of a galaxy: two each
for angular momentum vectors and its projections. We count all
these possibilities independently in the analysis.
%__________________________________________________________________
\subsection{Expected distribution: numerical simulation}

Aryal \& Saurer (2000) performed random simulations imposing
various types of selections in the database and concluded that any
selections can cause the changes in the expected isotropic
distribution curves for both angular momentum vectors (polar
angles) and its projections (azimuthal angles). We noticed
following selection effects in our database: (1) the positions of
galaxies in the clusters are inhomogeneous. (2) the PAs of face-on
($i$$\sim$ 0$^{\circ}$) are mostly unknown and (3) the deficiency
of high inclination ($i$$\sim$ 0$^{\circ}$) galaxies. We perform
random simulation method proposed by Aryal \& Saurer (2000) in
order to find expected isotropic distribution curves ($\theta$ and
$\phi$) by removing above mentioned selection effects. We apply
cosmological principle by assuming isotropic distribution of
angular momentum vectors of 10$^{7}$ virtual galaxies and use the
formulae (1) and (2) in the random simulation. Therefore, the
inclination angle ($i$) and latitude ($B$) are distributed as
$\propto$ $sin\ i$ and $\propto$ $cos\ B$, respectively and the
variables longitude ($L$) and position angle (PA) are distributed
randomly (Aryal \& Saurer 2000).

\subsection{Statistical tests}

We perform chi-square, autocorrelation, Fourier (Godlowski 1993),
Kolmogorov-Smirnov (K-S) (Press et al. 1992, Kanji 1995) and
Kuiper-V (Kuiper 1962) tests in order in the observed and expected
distributions to discriminate anisotropy from isotropy. The
details about these statistical tests are given in the appendix of
Aryal et al. (2007). The limits for anisotropy are as follows:
\begin{itemize}
\item chi-square probability (P$(>\chi^2)$) $<$ 0.050,

\item auto-correlation coefficient (C/C($\sigma$)) $>$ 1,

\item first order Fourier coefficient
($\Delta_{11}$/$\sigma$($\Delta_{11}$)) $>$ 1,

\item Fourier probability (P($>\Delta_1$)) $<$ 0.150,

\item K-S = 1  \item Kuiper-V = 1.

\end{itemize}
In last two statistical tests, null hypothesis (isotropy) is
represented by ``0" that can not be rejected at the chosen
significance level whereas the value ``1" designates that the null
hypothesis can be rejected (anisotropy).

The first order Fourier coefficient ($\Delta_{11}$) provides
information regarding preferred alignments. A positive (negative)
value of first order Fourier coefficient ($\Delta_{11}$) in the
$\theta$-distribution suggests that the angular momentum vectors
of galaxies tend to orient parallel (perpendicular) with respect
to the reference coordinate system. Similarly, a positive
(negative) $\Delta_{11}$ in the $\phi$-distribution indicates that
the projections of angular momentum vectors of galaxies in the
galactic (G) or supergalactic (S) planes tend to point radially
(tangentially) towards the center of the reference coordinate
system.
%________________________________________________________
\section{Results}
Table 2 shows the values of statistical parameters with respect to
galactic (G) and supergalactic (S) coordinate systems. In this
section we discuss polar ($\theta$) and azimuthal ($\phi$) angle
distributions of galaxies in each clusters separately.
%-------------------------------------------------------------------------------------
\begin{table*}
\caption[]{Statistical values for both $\theta$ and $\phi$
distributions. The Abell number of the cluster, values of
chi-square probability (P$(>\chi^2)$) and auto-correlation
coefficients (C/C($\sigma$)) are given in first three columns. The
fourth-seventh columns list the values of first order Fourier
coefficient $\Delta_{11}$/$\sigma$($\Delta_{11}$), first order
Fourier probability P($>\Delta_1$), results of Kolmogorov-Smirnov
(K-S) and Kuiper-V (KV) tests, respectively. }
$$
 \begin{array}{p{0.13\linewidth}ccccccc}
            \hline
            \noalign{\smallskip}
            Abell  &  $P$(>\chi^2)$$ & $C/C($\sigma$)$  &   $$\Delta_{11}$/$\sigma$($\Delta_{11}$)$  &  $P($>\Delta_1$)$ & $K-S$   &  $Kuiper-V$  \\
                   &  $G$/$S$        &   $G$/$S$        &  $G$/$S$                                   & $G$/$S$           & $G$/$S$ &  $G$/$S$  \\
                   \noalign{\smallskip}
            \hline
            \noalign{\smallskip}
            polar angle \\
            \hline
            S1171   &   0.959   /   0.820   &   -0.5/+0.1  &   +0.0/-0.6    &   0.999   /   0.805   &   0/0 &   0/0 \\
            S0001   &   0.554   /   0.779   &   -0.4/+0.7  &   -0.6/+1.2    &   0.798   /   0.435   &   0/0 &   0/0 \\
            1035    &   0.721   /   0.795   &   +0.1/-1.0  &   +0.4/-0.5    &   0.864   /   0.868   &   0/0 &   0/0 \\
            1373    &   0.495   /   0.553   &   +0.7/-0.9  &   -0.2/-0.2    &   0.796   /   0.964   &   0/0 &   0/0 \\
            1474    &   0.012   /   0.610   &   -1.8/+0.2  &   +1.1/-0.9    &   0.380   /   0.501   &   0/0 &   0/0 \\
            4053    &   0.718   /   0.937   &   -1.3/+0.1  &   -0.4/-0.6    &   0.900   /   0.825   &   0/0 &   0/0 \\
            \hline
            azimuthal angle\\
            \hline
            S1171   &   0.264   /   0.759   &   +0.1/-0.2  &   +2.0/-0.4    &   0.141   /   0.903   &   0/0 &   0/0 \\
            S0001   &   0.267   /   0.914   &   -0.8/-0.1  &   -1.8/+0.3    &   0.185   /   0.928   &   0/0 &   0/0 \\
            1035    &   0.851   /   0.154   &   +0.4/-1.7  &   -1.5/-0.2    &   0.296   /   0.883   &   0/0 &   0/0 \\
            1373    &   0.041   /   0.964   &   +0.7/0.1   &   -2.0/+0.1    &   0.019   /   0.980   &   0/0 &   0/0 \\
            1474    &   0.000   /   0.271   &   -0.7/-0.2  &   +0.9/+0.6    &   0.118   /   0.646   &   0/0 &   0/0 \\
            4053    &   0.064   /   0.104   &   -0.6/-0.7  &   -2.2/-1.2    &   0.095   /   0.333   &   0/0 &   0/0 \\
            \noalign{\smallskip}
            \hline
         \end{array}
     $$
\end{table*}
%-----------------------------------------------------------
\subsection{Abell S1171}
%------------------------------------------------------------
\begin{figure}
\centering \vspace{0.0cm}
     \centering \vspace{0.0cm}
      \includegraphics[height=3.6cm]{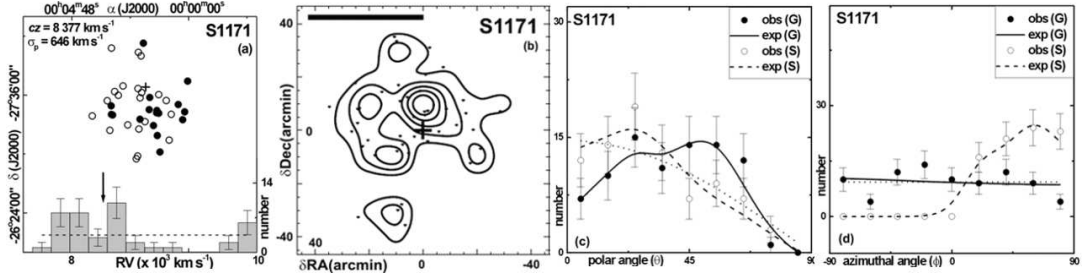}
      \caption[]{(a) Distribution of galaxies in Abell S1171. The galaxies having radial velocity ($cz$) $<$ ($>$) $\overline{cz}$ of the cluster is represented by solid (hollow)
      circle. The dashed line and an arrow represent the average distribution of $cz$ of galaxies and the cluster $\overline{cz}$.
      (b) Number density map: the member galaxies are represented by dots and the number density contours are overlaid.
      The cross and the thick horizontal bar indicate the cluster center and the the physical size of 1 Mpc.
      The polar ($\theta$) (c) and azimuthal ($\phi$) angle (d) distributions of galaxies with respect to the G- (solid line) and S- (dashed line) co-ordinate systems.
      The cosine and average distributions (dotted line) are shown for the comparison. The statistical $\pm1\sigma$ error bars are shown for the observed counts.}
\end{figure}

Abell S1171 is the spherically shaped nearby ($\overline{cz}$ =
8\,377 km s$^{-1}$) cluster in our database (Table 1). HL noticed
that a few galaxies have a large velocity deviations ($cz$ --
$\overline{cz}$ $\sim$1\,400 km s$^{-1}$) from the cluster main
body ($cz$ $\sim$8\,400 km s$^{-1}$). The position and radial
velocity ($cz$) distributions of member galaxies show a bimodal
velocity distribution (Fig. 1a). A large number of substructures
with central condensation can be seen in the number density map
(Fig. 2b).

Figure 1c,d shows the polar ($\theta$) and azimuthal angle
($\phi$) distributions of member galaxies in S1171. The solid and
dashed curves represent the expected isotropic distribution curves
obtained from random simulation. The polar angle, $\theta$ =
0$^{\circ}$ (90$^{\circ}$) corresponds to the angular momentum
vector tends to lie parallel (perpendicular) to the
galactic/Supergalactic plane. All six statistical parameters
suggest isotropy in the $\theta$-distribution (see Table 2). No
preferred alignment of angular momentum vectors of galaxies is
observed with respect to the G- and S-coordinate systems.
Therefore, a random orientation of angular momentum vectors of
galaxies is noticed in the cluster.

In the $\phi$-distribution, 0$^{\circ}$ corresponds to the
projections of angular momentum vectors tend to point radially
towards centre of the reference coordinate system (centre of the
Milky way in G-system and Virgo cluster center in S-system). No
humps and dips are observed in Fig. 1d, suggesting no preferred
alignments. It can be concluded that the galaxies with a large
velocity dispersion cause the cluster dynamically unstable as
discussed by Godlowski (2011b).

%-----------------------------------------------------------
\subsection{Abell S0001}
%------------------------------------------------------------
HL noticed substructures in 2D and 3D number density maps. A
low-velocity tail at $cz$ $\sim$ 7\,800 km s$^{-1}$ which is far
from the mean radial velocity of the cluster can be seen in Fig.
2a. Fig. 2b shows two widely separated local minima, strongly
suggests that the cluster might not be in the dynamical
equilibrium.
\begin{figure}
\centering \vspace{0.0cm}
     \centering \vspace{0.0cm}
       \includegraphics[height=3.6cm]{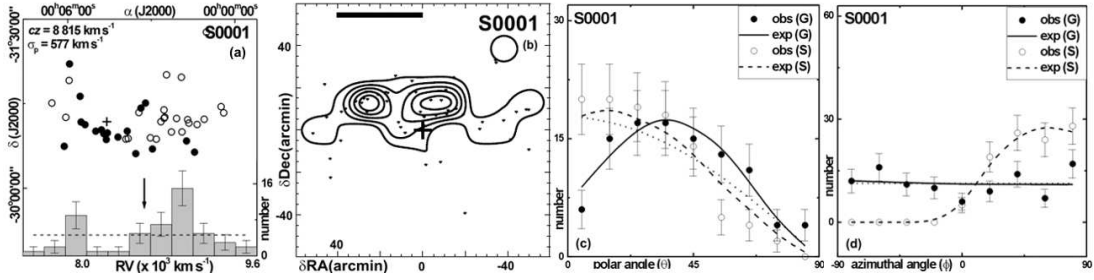}
      \caption[]{The cluster Abell S0001: the scatter plot (a), number density map (b), $\theta$ and (c)
      $\phi$ (d) distributions of galaxies. Symbols and explanations as in Fig. 1.}
\end{figure}
All six statistical parameters show isotropy in both the $\theta$-
and $\phi$-distributions (Table 2). A very good correlation
between the expected and observed distribution can be seen in the
histograms (Fig. 2c,d), suggesting a random orientation of angular
momentum vectors of galaxies. No preferred alignments of angular
momentum vectors of galaxies in the cluster Abell S0001 is found.
%________________________________________________________________________

%-----------------------------------------------------------
\subsection{Abell 1035}
%------------------------------------------------------------
This cluster has the largest velocity dispersion with two
subclusterings along north and south (Fig. 3a,b). HL found
substructures in all maps (1D, 2D and 3D), suggesting a large but
unequal velocity dispersion causing dynamically unstable. Using
photometric database (Lauberts 1982), Aryal and Saurer (2006)
studied the preferred alignments in this cluster and found that
the angular momentum vector of galaxies tend to lie in the Local
Supercluster (LSC) plane, supporting pancake model (Doroshkevich
1973) of galaxy formation. HL found that the angle between the
rotation axes and the LSC plane is about 56$^{\circ}$, suggesting
a bimodal distribution.
\begin{figure}
\centering \vspace{0.0cm}
     \centering \vspace{0.0cm}
      \includegraphics[height=3.6cm]{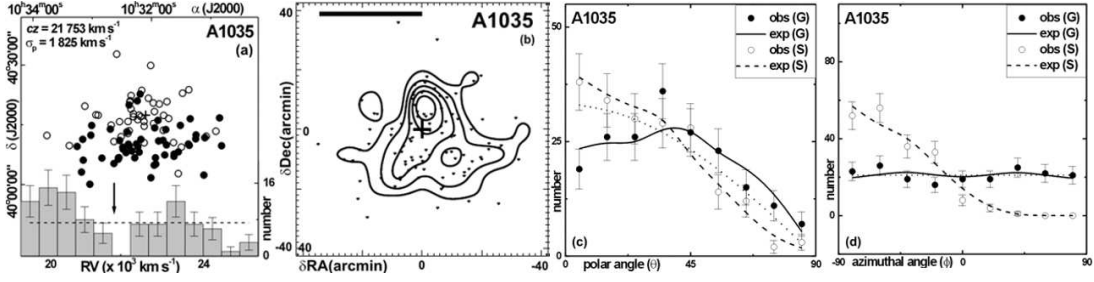}
      \caption[]{The cluster Abell A1035: the scatter plot (a), number density map (b), $\theta$ and (c)
      $\phi$ (d) distributions of galaxies. Symbols and explanations as in Fig. 1.}
\end{figure}
In the present study we used spectroscopic database (SDSS and
2dFGRS) and verified the prediction made by HL by observing a
significant hump at 35$^{\circ}$ ($>$1$\sigma$) in the
$\theta$-distribution (Fig. 3c), supporting bimodal distribution:
angular momentum vectors of galaxies tend to lie both parallel and
perpendicular with respect to the plane of the Milky way. In the
$\phi$-distribution, isotropy is noticed (Fig. 3d), suggesting
that the choice of co-ordinate system is independent of preferred
alignments.
%________________________________________________________________________

%-----------------------------------------------------------
\subsection{Abell 1373}
%------------------------------------------------------------
This cluster is the most distant ($\overline{cz}$ = 37\,595 km
s$^{-1}$) with bimodal velocity distribution spatially separated
by the rotation axes (Fig. 4a). The number density map is similar
to that of the cluster Abell 1035, i.e., substructure is seen in
1D-3D (HL).
\begin{figure}
\centering \vspace{0.0cm}
     \centering \vspace{0.0cm}
       \includegraphics[height=3.6cm]{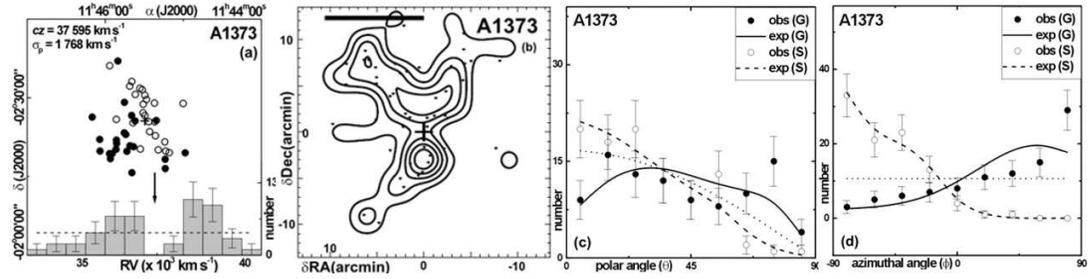}
      \caption[]{The cluster Abell A1373: the scatter plot (a), number density map (b), $\theta$ and (c)
      $\phi$ (d) distributions of galaxies. Symbols and explanations as in Fig. 1.}
\end{figure}
All statistical tests suggest isotropy in both the $\theta$- and
$\phi$-distributions, advocating hierarchy model (Peebles 1969) of
galaxy evolution. In the $\theta$-distribution, a significant hump
at 75$^{\circ}$ ($>$1.5$\sigma$) can be seen (Fig. 4c). In
addition, a dip at 45$^{\circ}$ ($\sim$1.5$\sigma$) is followed by
the hump. Therefore, a local tidal effect, probably due to the
substructure formation, is noticed. In the $\phi$-distribution, a
hump at 85$^{\circ}$ (($>$2$\sigma$) and dips at 65$^{\circ}$
($\sim$1.5$\sigma$) and 75$^{\circ}$ ($\sim$1.5$\sigma$) supports
local effect (Fig. 4d).
%________________________________________________________________________

%-----------------------------------------------------------
\subsection{Abell 1474}
%------------------------------------------------------------
Einasto et al. (2001) studied Virgo-Coma supercluster and
concluded that Abell 1474 is the member cluster of of that
supercluster. HL noticed three substructures in the number density
map (Fig. 5b), suggesting dynamically unstable rotating cluster
which is not in the dynamical equilibrium because of large value
of $\overline{cz}$--$\sigma_{v}$.
\begin{figure}
\centering \vspace{0.0cm}
     \centering \vspace{0.0cm}
      \includegraphics[height=3.6cm]{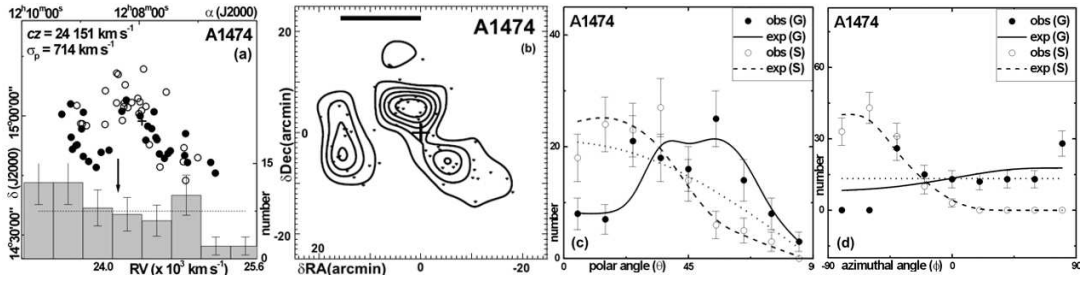}
      \caption[]{The cluster Abell A1474: the scatter plot (a), number density map (b), $\theta$ and (c)
      $\phi$ (d) distributions of galaxies. Symbols and explanations as in Fig. 1.}
\end{figure}
The chi-square and autocorrelation tests show anisotropy in the
$\theta$-distribution, whereas isotropy is noticed in the Fourier,
KS and Kuiper V tests. A hump at 25$^{\circ}$ ($\sim$2$\sigma$)
and a dip at 45$^{\circ}$ (1$\sigma$) suggest a local tidal effect
because of the subsclustering (Fig. 5c). In the
$\phi$-distribution, significant humps at $-$45$^{\circ}$
(2$\sigma$) and 90$^{\circ}$ (1.5$\sigma$) supports it (Fig. 5d).
Therefore, the spatial orientation of galaxies in the cluster
Abell 1474 shows a weak preference in the alignments.
%________________________________________________________________________

%-----------------------------------------------------------
\subsection{Abell 4053}
%------------------------------------------------------------
Porter \& Raychaudhury (2005) reported that the cluster A4053 is a
member clusters of Pisces-Cetus supercluster. HL found that the
cluster shows substructures in 1D and 3D maps. The number density
map (Fig. 6b) shows elongation along north-east and south-west
direction.
\begin{figure}
\centering \vspace{0.0cm}
     \centering \vspace{0.0cm}
       \includegraphics[height=3.6cm]{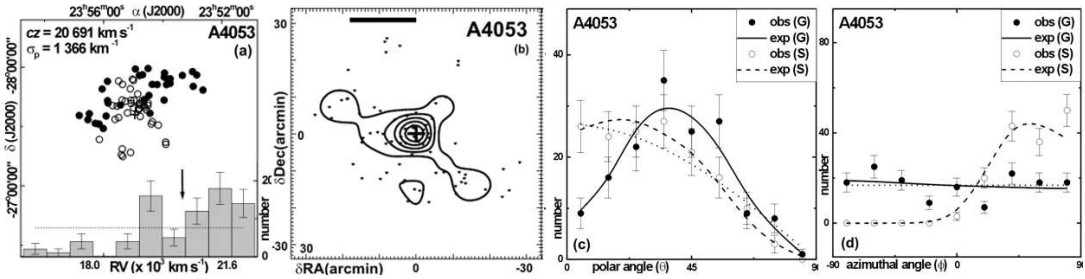}
      \caption[]{The cluster Abell A4053: the scatter plot (a), number density map (b), $\theta$ and (c)
      $\phi$ (d) distributions of galaxies. Symbols and explanations as in Fig. 1.}
\end{figure}
In the histogram of polar ($\theta$) angle distribution, a good
agreement between the expected and observed distribution is
noticed, suggesting no preferred alignments (Fig. 6c). All
statistical tests support this. In the $\phi$-distribution, humps
($-$70$^{\circ}$, 50$^{\circ}$) and dips ($-$20$^{\circ}$,
20$^{\circ}$) at 1.5$\sigma$ level are significant (Fig. 6d),
suggesting anisotropy. Statistical tests support this result.
Therefore, projections of angular momentum vectors of galaxies in
A4053 tend to be oriented perpendicular towards the plane of the
Milky way, whereas no preference is noticed with respect to Virgo
cluster centre. Therefore a local effect can not be ruled out in
the cluster which has a multiple number-density peaks and large
value of velocity dispersion.
%________________________________________________________________________

\section{Conclusion}
The preferred alignments of angular momentum vectors of galaxies
in six clusters having multiple number-density peaks with a
spatial segregation of high- and low-velocity galaxies are
studied. We adopted `position angle - inclination' method (Flin \&
Godlowski 1986) to compute three dimensional parameters (polar and
azimuthal angles of the galaxy rotation axes) using
two-dimensional observed parameters (e.g., positions, diameters,
position angles). To remove selection effects from the database, a
numerical simulation is performed, as proposed by Aryal \& Saurer
(2000). The observed and expected isotropic distributions are
compared using five statistical tests namely, chi-square,
auto-correlation, Fourier, K-S and Kuiper V.

In general, no preferred alignments is noticed for all six
clusters supporting hierarchy model as predicted by Peeples
(1969). However, local effects are noticed in the clusters that
have substructures in 1D, 2D and 3D analysis (HL). Therefore, a
large value of velocity dispersion with substructures in the
clusters do not lead their galaxies to support pancake
(Doroshkevich 1973) and primordial vorticity theory (Ozernoy
1978). A very good correlation between the hierarchy (Peebles
1969) and Li model (1998) is found, as in our previous work (Aryal
et al. 2013). Therefore, vanishing angular momenta favor the
formation of substructures in the clusters that have large
velocity dispersion. The preferred alignment is found to increase
with the cluster richness. Therefore it can be interpreted as an
effect of tidal forces mechanism (Heavens \& Peacock 1988, Catela
\& Theuns 1996, Stephanovich \& Godlowski, 2015), but also is in
agreement with Li's (1998) model in which galaxies form in the
rotating universe.

The tidal torque naturally arises in the hierarchical clustering
scenario and hence the distribution of angular momentum vectors of
galaxies becomes random. However, a tidal torque shear tensor (due
to gravitational effect) can cause a local preference on angular
momentum vectors as predicted by Lee (2004) and Trujillo et al.
(2006).

%_____________________________________________________________

\begin{acknowledgements}
The authors thank the anonymous referee whose remarks contributed
to improve the paper. We acknowledge Dr. Ho Seong Hwang of
Department of Physics and Astronomy, Seoul National University,
Korea, for providing database. One of the authors (SNY)
acknowledges Central Department of Physics, Tribhuvan University,
Nepal for all kinds of support for his Ph.D. work.
\end{acknowledgements}

\label{lastpage}

\end{document}